\documentstyle[12pt,epsf]{article}

\newcommand{\Pdat}{June 1996}       
\newcommand{\Pnum}{96/20}           
\newcommand{\Ptit}{                 
Production of Nonlinear SUSY Higgs Bosons\\
at $e^+ e^-$ Colliders
}

\begin{document}
\begin{figure}[ht]
\vspace*{-1.5cm} 
  \hfill
\parbox[b]{4.0cm}{
\vbox to 22mm {
{\footnotesize 
  \begin{flushright}
   \raisebox{1.3mm}{\large \sf PITHA \Pnum}\\
    \rule[0.2cm]{4.0cm}{1.0mm}\\
   {\large \sf \Pdat }\\
  \end{flushright}
}
}
}
\end{figure}

\vfill

\setcounter{footnote}{-1}

\begin{center} \def\thefootnote{*}

   {\LARGE \sf \Ptit
\footnote{
To be published in Physics Letters B. --- This work is supported in part by the Ministery of Education, Korea, BSRI-96-2442.}

}
\vfill  
{\large\sf 
            S.W.~Ham$^1$, H.~Genten$^2$, B.R.~Kim$^2$, S.K.~Oh$^1$}
\bigskip

{\normalsize\sf %
$^1$ Department of Physics, Kon-Kuk University, Seoul, Korea}\\
{\normalsize\sf %
$^2$ III. Physikalisches Institut, RWTH Aachen, Germany}\\
\end{center}

%

\vfill

\begin{abstract}
We investigate the Higgs sector of a nonlinear supersymmetric standard
model at LEP 1 and LEP 2, as well as at future linear $e^+e^-$ colliders
with $\sqrt{s}$ = 500, 1000, and 2000 GeV.
The LEP 1 data do not put any constraints on the parameters of the model,
and allow a massless Higgs boson in particular.
For LEP 2, there are remarkable differences between the Higgs productions
at $\sqrt{s}$ = 175 GeV on the one hand and that at $\sqrt{s}$ = 192 GeV
and 205 GeV on the other hand.
The case for $\sqrt{s}$ = 175 GeV is similar to LEP 1, whereas those for
$\sqrt{s}$ = 192 GeV and 205 GeV will be able to give experimental
constraints on the parameters.
Finally the $e^+e^-$ colliders with $\sqrt{s}$ = 500, 1000, and 2000 GeV are most
probably able to test the model conclusively.
\end{abstract}

\vspace*{\fill}

\newpage

\section{Introduction}

For more than a decade the phenomenology of supersymmetric models has
been studied, and the search for supersymmetric particles is one of
the main goals of 
existing and future accelerators.
Most of the 
supersymmetric models investigated so far are 
linear ones, i.e., 
supersymmetry is realized linearly in them \cite{nls1}.
However, it is still an open question whether 
supersymmetry is
reali\-zed linearly or nonlinearly.

The formalism for extending the standard model 
non\-linear-super\-sym\-me\-tri\-cally was developed by Samuel and Wess \cite{nls2}.
Recently one of us has constructed the general form of a nonlinear supersymmetric standard
model in curved space and derived the Higgs potential in the flat
limit \cite{nls3}.
In global nonlinear supersymmetric models the only new particle is
the Akulov-Volkov field \cite{nls4}, which is a Goldstone fermion.
Experimentally, no Goldstino has been observed.
In local nonlinear supersymmetric models this goldstino can be gauged away;
it is absorbed into the gravitino, which becomes massive \cite{nls5}.
In the flat limit, the supergravity multiplet decouples from the
ordinary matter 
with the only reminiscence of supersymmetry manifesting itself in the Higgs 
sector.

The Higgs sector of the nonlinear SUSY models is evidently larger 
than that of the Standard model.  It contains at least two dynamical Higgs 
doublets and an auxiliary Higgs singlet.  In the case that both a dynamical 
and an auxiliary singlet are included in the theory, the Higgs boson spectrum 
of the nonlinear model resemble that of the linear next-to-minimal 
supersymmetric standard model (NMSSM).  In both models, there are three scalar
Higgs bosons, two pseudoscalar Higgs bosons, and a pair of charged Higgs
bosons.  However, the structure of the Higgs potential is different between
nonlinear and linear supersymmetry.

In this article we investigate the phenomenology of the nonlinear 
supersymmetric model with both an auxiliary and a dynamical Higgs singlet 
beside the doublets.  In particular, we are interested in how far the Higgs sector can be tested at LEP, as well as at future $e^+ e^-$ colliders.

\section{The Model}
\label{nlsmodel}

The complete Higgs potential of our model is given in Ref.\ \cite{nls3}: 
\begin{eqnarray}
        V &=& {1\over 8}(g_1^2 +g_2^2)(|H^1|^2 -|H^2|^2)^2
                +{1\over2}g_2^2 |H^{1+} H^2|^2  \cr
        & & \mbox{ }+\mu^2_1 |H^1|^2 +\mu^2_2 |H^2|^2 +\mu^2_0 |N|^2  \cr
        & & \mbox{ }+\lambda_0^2 |H^{1T}\epsilon H^2|^2
                        +k^2 |N^{\dag} N|^2
                        +|N|^2 (\lambda^2_1 |H^1|^2 +\lambda^2_2 |H^2|^2) \\
        & & \mbox{ }+k\lambda_0 [(H^{1T}\epsilon H^2)^{\dag} N^2 
                        +{\rm h.c.}]  \cr
        & & \mbox{ }+[\lambda_1\mu_1|H^1|^2 N +\lambda_2\mu_2|H^2|^2 N
                        +\lambda_0\mu_0(H^{1T}\epsilon H^2)^{\dag} N
                        +{\rm h.c.}]  \cr
        & & \mbox{ }+[k\mu_0 N^{\dag} N^2 +{\rm h.c.}]  \ . \nonumber
\end{eqnarray}
The two Higgs doublets $H^1$ and $H^2$ and the singlet $N$ develope
vacuum expectation values $v_1$, $v_2$, and $x$, respectively.
The full mass matrices can be found in Ref.\ \cite{nls3}.
We denote the three scalar Higgs bosons and their masses respectively by
$S_1$, $S_2$, $S_3$ and $m_{S_1}$ $\le$ $m_{S_2}$ $\le$ $m_{S_3}$.
As in the case of the NMSSM \cite{nls7}, one can derive an upper bound on $m_{S_1}$
in our model as
\begin{equation}
        m^2_{S_1} \le m_{{S_1},{\rm max}}^2 =
                m_Z^2 \left(\cos^2 2\beta
                + {2\lambda_0^2 \over g_1^2 + g_2^2}\sin^2 2\beta \right) \ ,
\label{ms1bound}
\end{equation}
where $\tan\beta = v_2/v_1$.
This relation shows that the quartic coupling $\lambda_0$ is relevant
for the upper bound as in the case of the standard model.
For $\lambda_0^2 \le (g_1^2 + g_2^2)/2 = (0.52)^2$ this relation gives
$m_{S_1}^2 \le m_Z^2$, whereas for $\lambda_0^2 > (0.52)^2$ the upper
bound is given by $m_{S_1}^2 \le (1.92 \lambda_0 m_Z)^2$.
In the latter case, the upper bound of $\lambda_0$ determines that of
$m_{S_1}$.
For $m_t = 175$ GeV (190 GeV) and with the GUT scale as the cut-off scale,
one obtains $\lambda_{0,{\rm max}} \approx$ 0.74 (0.66) and
$m_{S_1} \le 130$ GeV (117 GeV).

It is very instructive to notice that, as in the case of the NMSSM \cite{nls8},
the upper bounds on $m_{S_2}$ and $m_{S_3}$ can be derived as functions
of $m_{S_1,{\rm max}}$ and $m_{S_1}$:
\begin{eqnarray}
        & & m^2_{S_2} \le m^2_{{S_2},{\rm max}}
                = {m^2_{{S_1},{\rm max}} -R^2_1 m^2_{S_1} \over 1 - R^2_1} \cr
        & & m^2_{S_3} \le m^2_{{S_3},{\rm max}}
                = {m_{{S_1},{\rm max}}^2 -(R^2_1 +R^2_2) m^2_{S_1}
                        \over 1 - (R^2_1 +R^2_2)}
\end{eqnarray}
with
\begin{eqnarray}
        & & R_1 = U_{11}\cos\beta + U_{12} \sin\beta \cr
        & & R_2 = U_{21}\cos\beta + U_{22} \sin\beta    \ ,
\end{eqnarray}
where $U_{ij}$ is the orthogonal matrix that diagonalises the scalar
mass matrix, and $0 \le R^2_1 +R^2_2 \le 1$.
Clearly, $R_1$ and $R_2$ are complicated functions of the relevant paremeters.
Nevertheless these relations turn out to be very useful to derive the lower
limits on the production cross section of the scalar Higgs bosons.

\section{Higgs Production at LEP 1 and LEP 2}

The LEP 1 data yield an experimental lower bound of 60 GeV on the Higgs
boson mass of the standard model, and 44 GeV for the mass of the lightest
scalar Higgs boson of the minimal linear supersymmetric standard model (MSSM).
In the case of the NMSSM the LEP 1 data do not exclude the existence of a
massless scalar Higgs boson \cite{nls9}.
Now, we analyse the LEP 1 data in the frame of our 
model.
As in the case of the NMSSM, the main contributions to the production cross
section come from (i) the Higgs-strahlung process, (ii) the process
where $S_i$ is radiated off leptons or quarks, and (iii) associated pair
production $P_j S_i$, where $P_j$ $(j=1,2)$ is a pseudoscalar Higgs boson:
\begin{equation}
\begin{array}{ccccc}
({\rm i})   & Z & \ \ \rightarrow Z^* S_i  & \rightarrow & {\bar f} f S_i  \cr
({\rm ii}) & Z & \rightarrow f {\bar f} & \rightarrow & {\bar f} f S_i   \\
({\rm iii})  & Z & \ \ \rightarrow P_j S_i  & \rightarrow & {\bar f} f S_i  \ ,
\nonumber
\end{array}
\label{processes}
\end{equation}
The dominant contributions come from the $b$ quark, $f = b$.

\begin{figure}
\epsfxsize=6.7cm \hspace*{3.75cm}
\epsffile{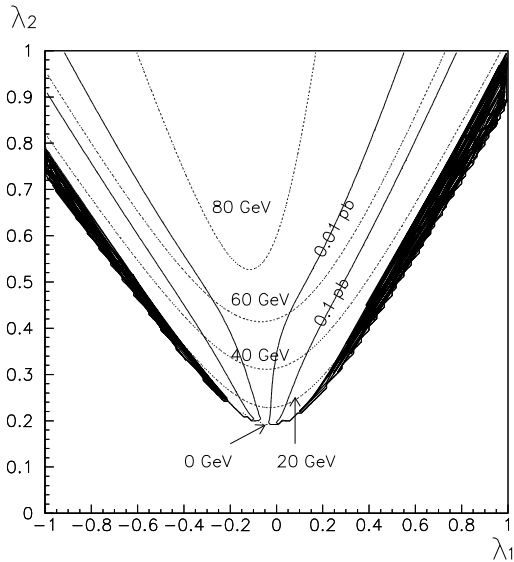}
\caption[plot]{Contour lines of the lightest Higgs boson mass $m_{S_1}$
(dotted) and of the production cross section $\sigma_1$ (solid) at $\sqrt{s}
= m_Z$, as functions of $\lambda_1$ and $\lambda_2$, for $\lambda_0$ = 0.4,
$k = 0.02$, $\tan \beta$ = 3, and $m_C$ = 200 GeV. The shadowed region marks
the parameter region excluded by {\rm LEP} 1, defined as the region where the
production cross section is greater than 1 {\rm pb}.}
\label{nlsusy1}
\end{figure}

Our model has 6 free parameters which can be taken as $\lambda_0, k, \lambda_1,
\lambda_2, \tan \beta$ and $m_C$ (the charged Higgs mass).
We search for parameter regions where none of $S_i$ $(i=1,2,3)$ has enough
production cross sections to be detected at LEP 1 and where $m_{S_1}$ = 0
is still allowed.
Fig.\ \ref{nlsusy1} shows such a region. We plot for $\lambda_0$ = 0.4, $k$ = 0.02,
$m_C$ = 200 GeV, and $\tan \beta$ = 3 the contours of $m_{S_1}$ and the
production cross section of $S_1$, $\sigma_1$.
Only the shadowed region where $\sigma_1 \ge$ 1 pb is excluded by the LEP 1
data.
$\sigma_2$ is smaller than 12.3 fb and $\sigma_3$ vanishes in the entire plane.
Thus we conclude that the LEP 1 data allow the existence of massless $S_1$
in our model.

\begin{figure}
\epsfxsize=6.5cm
\mbox{\epsffile{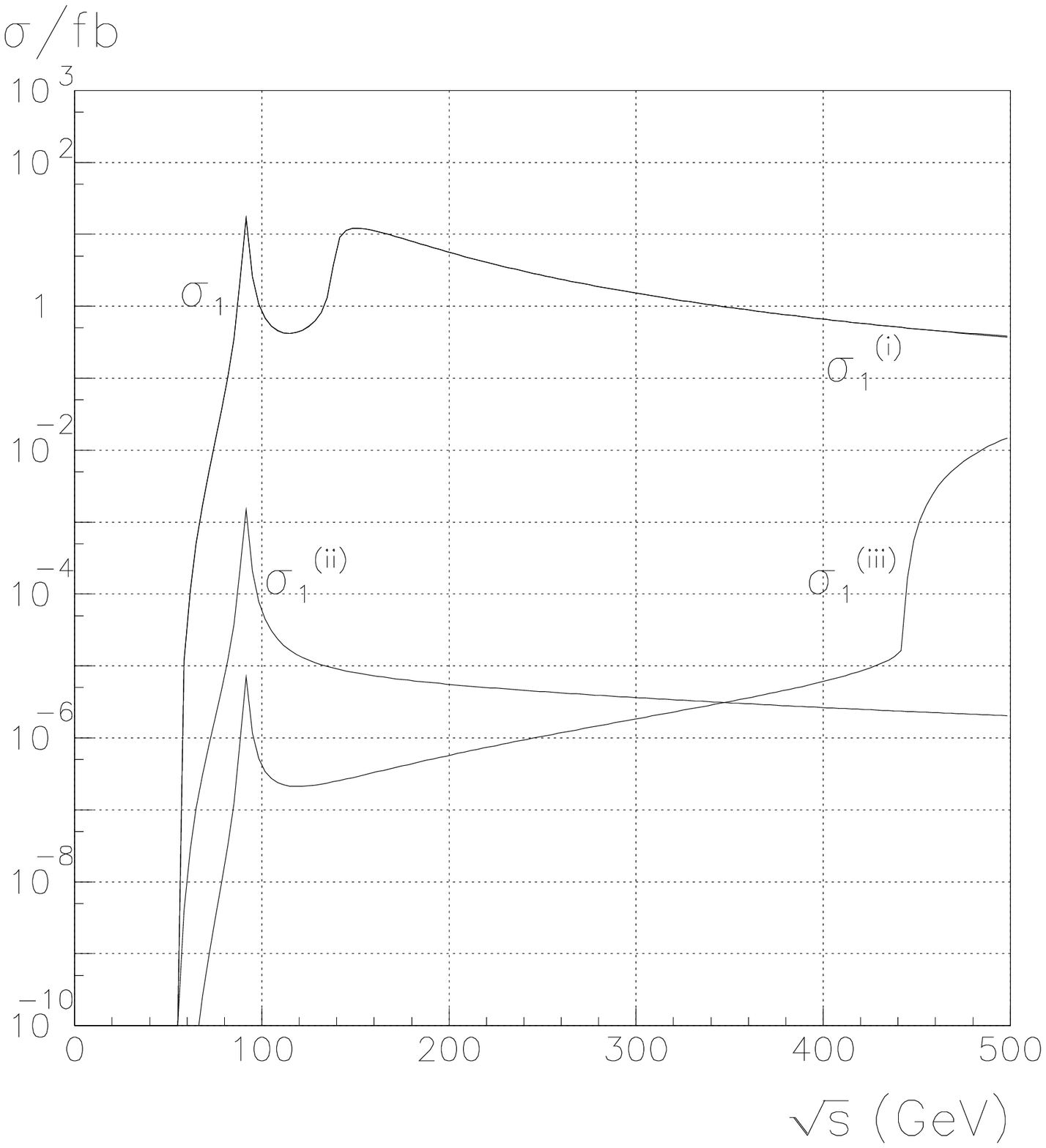}}
\hspace*{\fill} \epsfxsize=6.5cm
\mbox{\epsffile{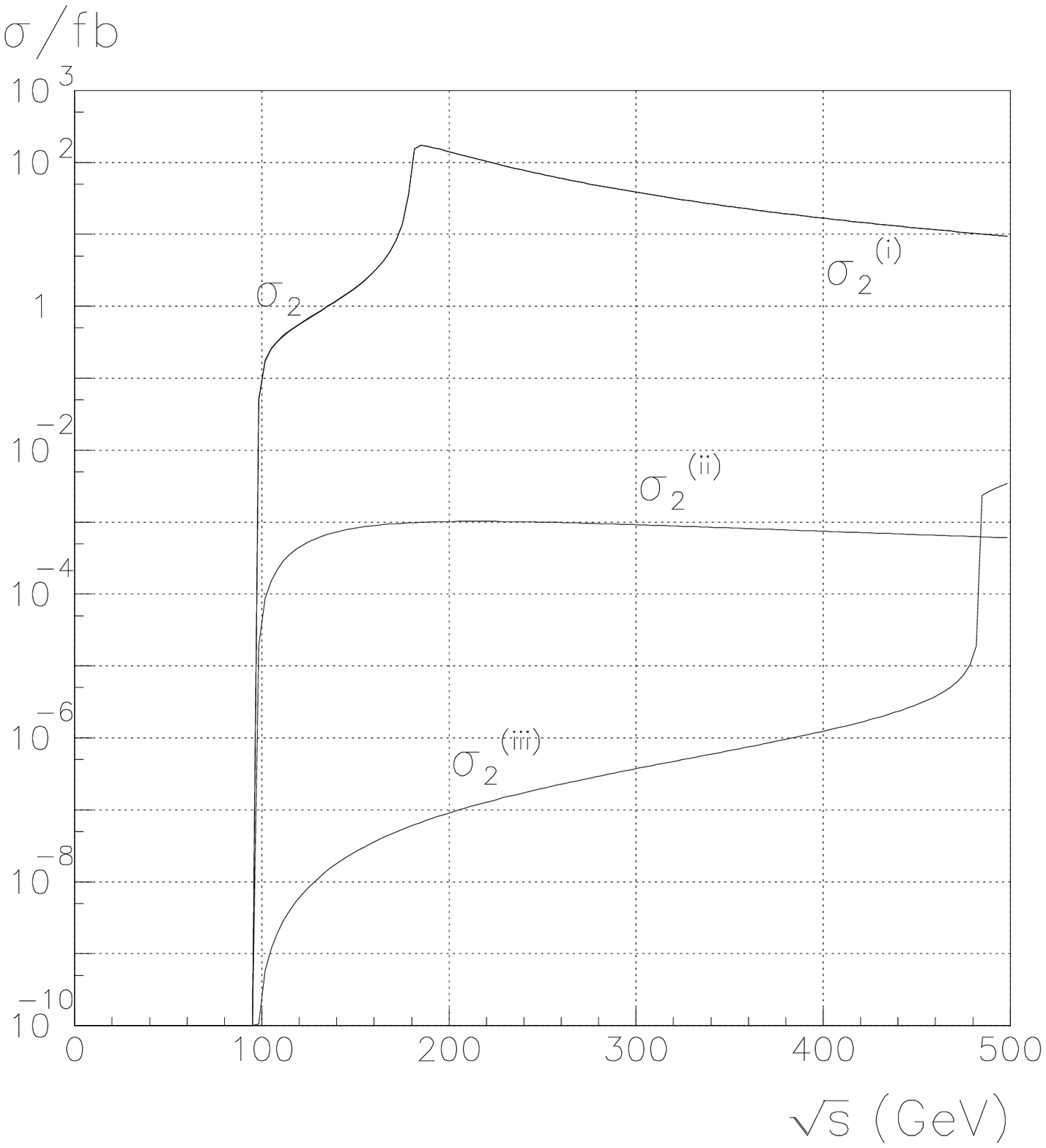}}
\\[-2cm] \hspace*{5.35cm} \large (a)
\\[-\baselineskip] \hspace*{13.45cm} \large (b) 
\\[2cm] \epsfxsize=6.5cm \hspace*{3.85cm}
\mbox{\epsffile{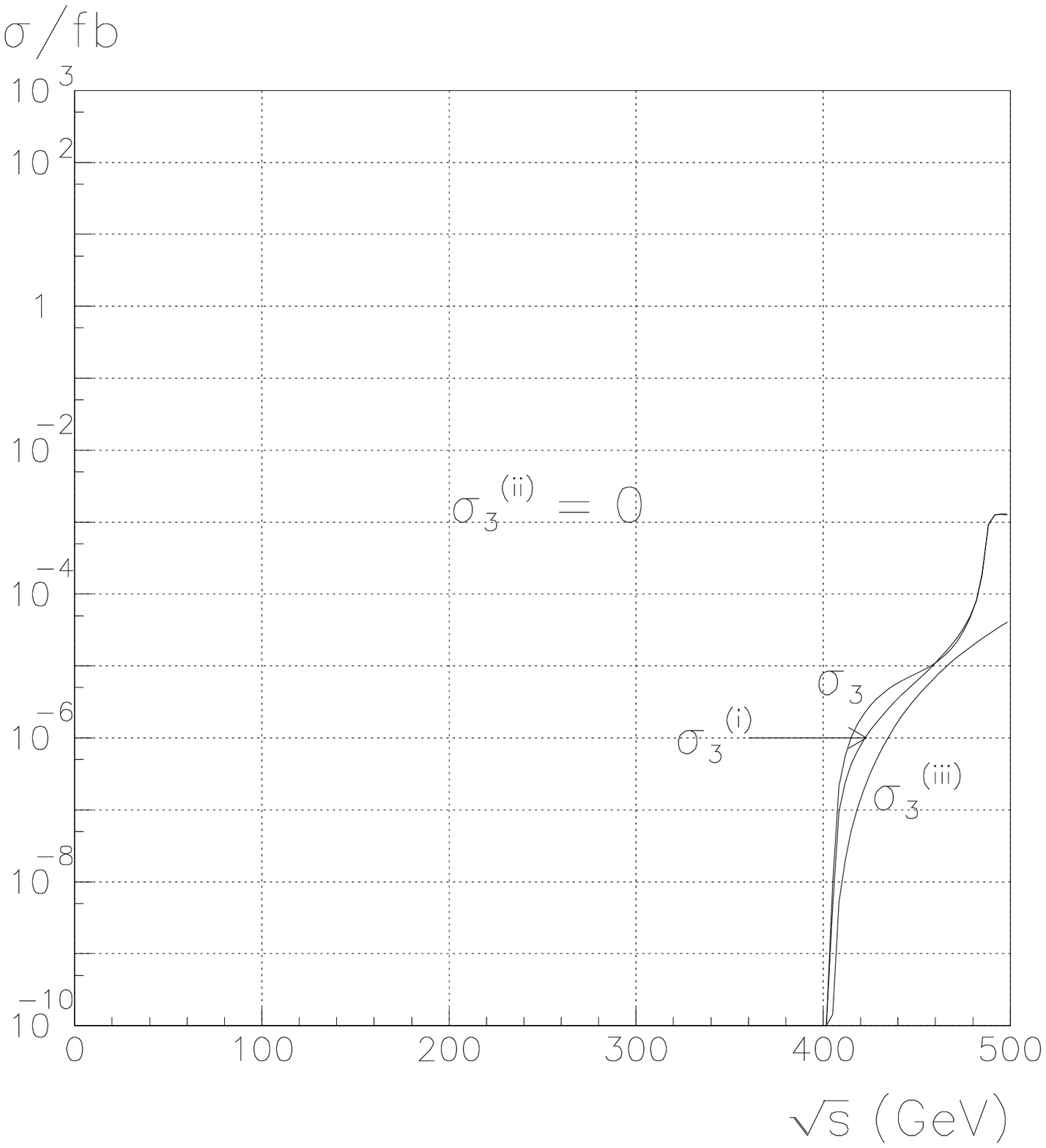}}
\\[-2cm] \hspace*{9.45cm} \large (c)
\\[2cm] \\[-2\baselineskip]
\caption[plot]{
(a) The production cross sections $\sigma_1^{\rm (a)}\ \rm (a=1,2,3)$ and
$\sigma_1$ of $S_1$ for 
$\lambda_0$ = 0.3, $k$ = 0.02, $\lambda_1$ = - 0.3, $\lambda_2$ = 0.45, 
$\tan \beta$ = 6, and $m_C$ = 400 GeV.
$\sigma_1^{\rm (a)}$ and $\sigma_1$ denote the production cross sections for the three processes of Eq.\ (\ref{processes}) and for the sum of these processes\ref{nlsusy1}, respectively.  $\sigma_1^{\rm (i)}$ is approximately equal to $\sigma_1$.\ \ ---\\
(b) The according production cross sections of $S_2$.  $\sigma_2^{\rm (i)}$
 is approximately equal to $\sigma_2$.\ \ ---\ \
(c) The according production cross sections of $S_3$.  $\sigma_3^{\rm (ii)}$ 
is neglegible.
}
\label{nlsusy2}
\end{figure}
Now we turn to LEP 2. In order to obtain a feeling we plot in Fig.\ 2a, 2b,
and 2c the production sections of $\sigma_1^{\rm (a)}$, $\sigma_2^{\rm (a)}$, 
and  $\sigma_3^{\rm (a)}$ $\rm (a = i,ii,iii)$ for the three processes of Eq.\ (\ref{processes}),
as well as the production cross sections $\sigma_1$, $\sigma_2$, and $\sigma_3$
for the sum of their processes, against the c.m.\ energy of the $e^+ e^-$ 
collider for a fixed set of parameters.  The contribution from the 
Higgs-strahlung process is dominant for $S_1$ and $S_2$.  $\sigma_1^{\rm (i)}$ 
and $\sigma_2^{\rm (i)}$ are approximately equal to $\sigma_1$ and $\sigma_2$,
with $\sigma_1 > \sigma_1^{\rm (i)}$ and $\sigma_2 > \sigma_2^{\rm (i)}$, 
respectively. The dominant contribution of interferences between three processes arises positively at the $Z$ peak for $S_1$, negatively at $\sqrt s \approx 180$ GeV for $S_2$, and negatively at $\sqrt s = 500$ GeV for $S_3$.  The maximum values of their contributions are $\sim 1$ fb for $S_1$, $\sim 10^{-1}$ fb for $S_2$, and $\sim 10^{-4}$ fb for $S_3$.
We observe that $\sigma_2$ is a very steep function of $\sqrt{s}$ in the
range between 150 GeV and 180 GeV, whereas it is rather flat
between $\sqrt{s}$ = 180 GeV and 240 GeV.
Thus we expect a relevant difference between the behavior 
at $\sqrt{s}$ = 175 GeV on the one hand and that at $\sqrt{s}$ = 192 GeV
or 205 GeV on the other hand.
It turns out that this expectation is generally correct when
the sum $(\sigma_1$ + $\sigma_2)$ is around 30 fb $\sim$ 50 fb.
The discovery
%
\begin{figure}
\epsfxsize=6.7cm \hspace*{3.75cm}
\epsffile{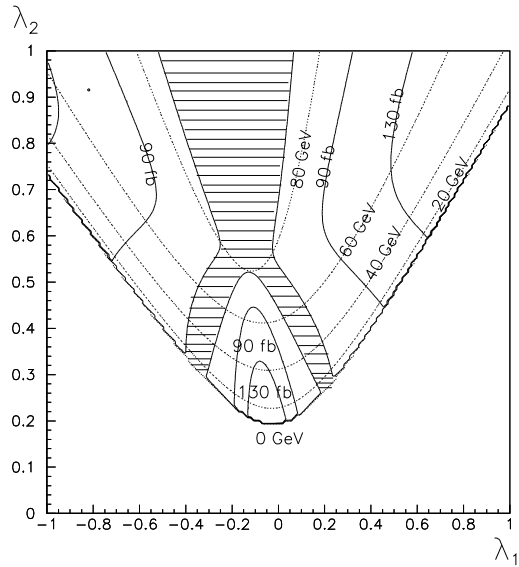}
\caption[plot]{Contour lines of the lightest Higgs boson mass $m_{S_1}$
(dotted) and of the production cross section $(\sigma_1 + \sigma_2)$
(solid) at $\sqrt{s}$ = 175 GeV, as functions of $\lambda_1$ and $\lambda_2$,
for $\lambda_0$ = 0.4, $k = 0.02$, $\tan \beta$ = 3, and $m_C$ = 200 GeV.
The cross section $(\sigma_1 + \sigma_2)$ is smaller than 50 fb in the
hatched region.}
\label{nlsusy3}
\end{figure}
%
\begin{figure}
\epsfxsize=6.5cm
\mbox{\epsffile{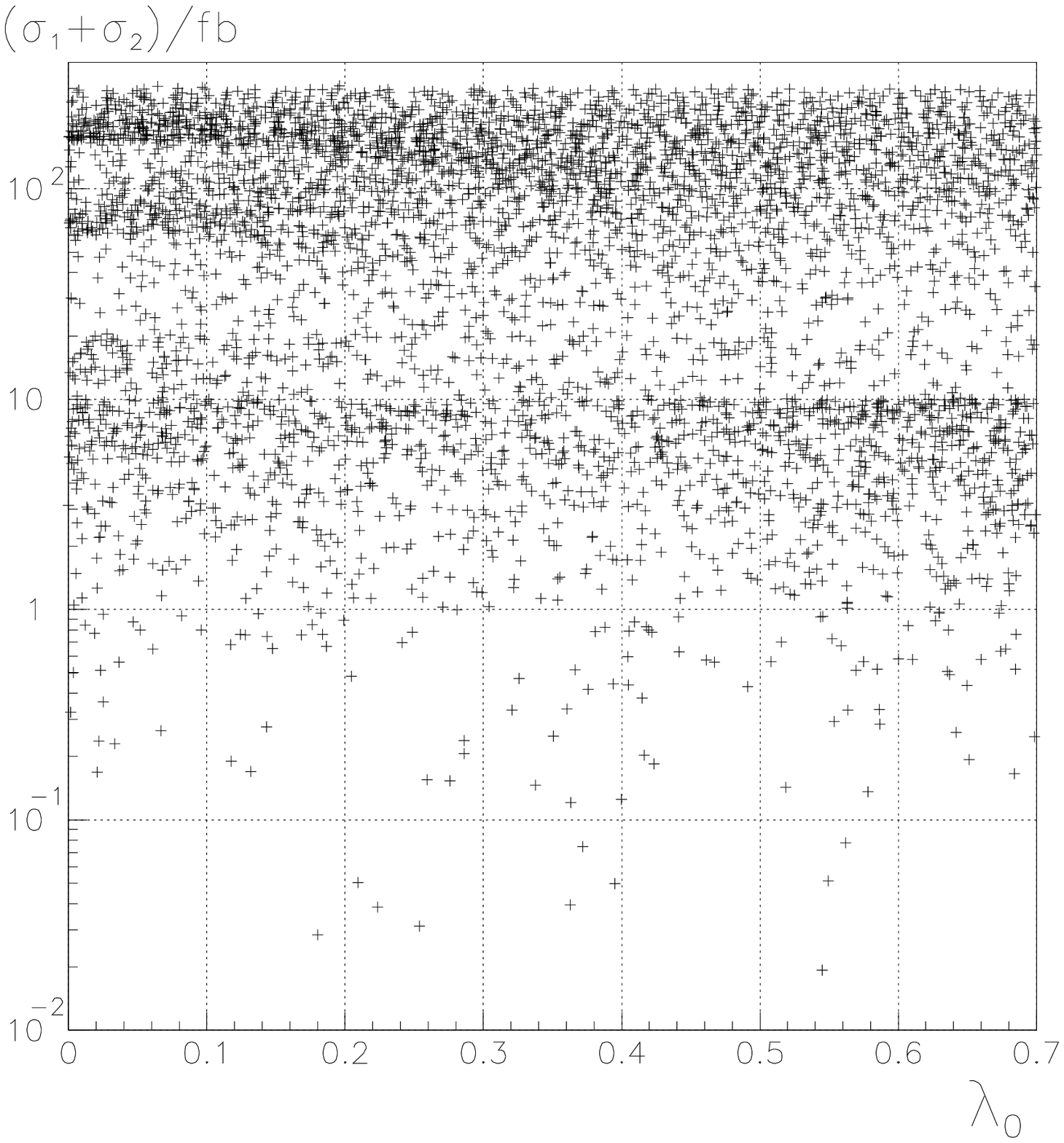}}
\hspace*{\fill} \epsfxsize=6.5cm
\mbox{\epsffile{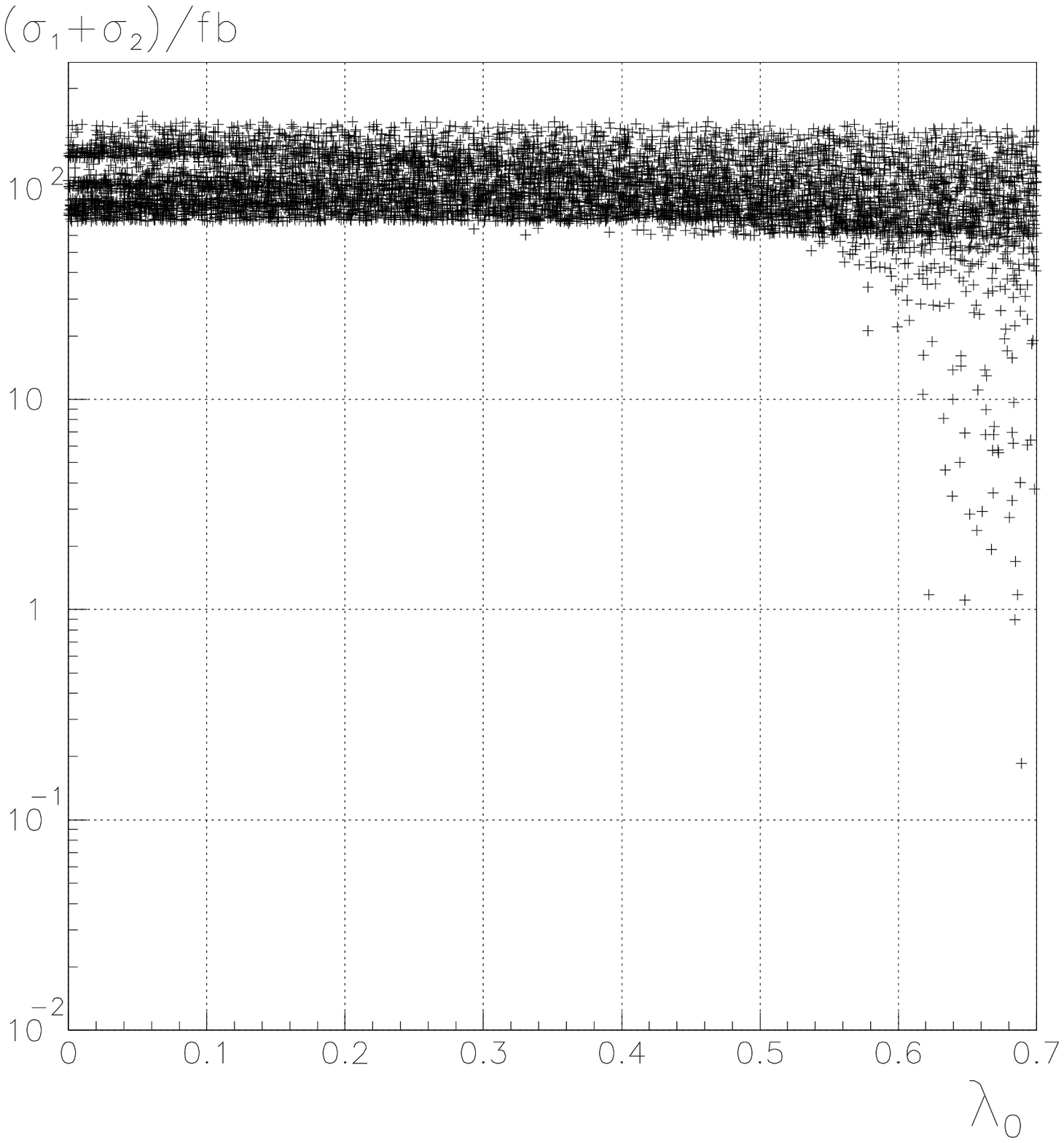}}
\\[-2cm] \hspace*{5.35cm} \large (a)
\\[-\baselineskip] \hspace*{13.45cm} \large (b)
\\[2cm] \epsfxsize=6.5cm \hspace*{3.85cm}
\mbox{\epsffile{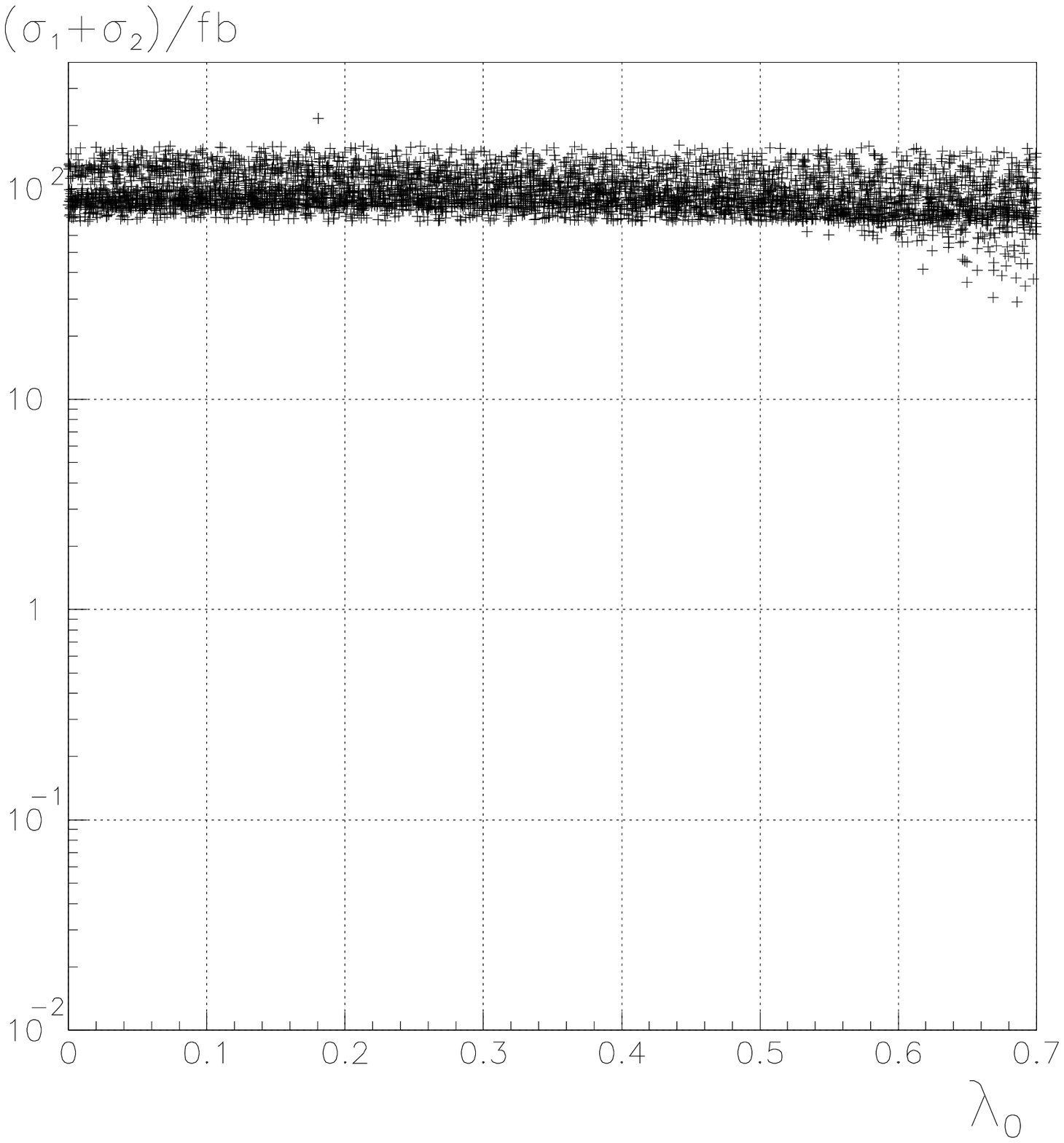}}
\\[-2cm] \hspace*{9.35cm} \large (c)
\\[2cm] \\[-2\baselineskip]
\caption[plot]{The cross section $(\sigma_1 + \sigma_2)$, as a function of 
$\lambda_0$, for $| k | \le 0.7$, $| \lambda_1 | \le 1$, $0 < \lambda_2 < 1$, 
$2 < \tan \beta < 15$, and 150 GeV $< m_C < 1000$ GeV, at (a) $\sqrt s = 175$ GeV, (b) $\sqrt s = 195$ GeV, (c) $\sqrt s = 205$ GeV.}
\label{nlsusy4}
\end{figure}
limit is about 50 fb for $m_S$ = 80 GeV and 30 fb for
$m_S$ = 40 GeV at a luminosity of 500 pb$^{-1}$ for $\sqrt{s}$ = 175 GeV and
at one of 300 pb$^{-1}$ for $\sqrt{s}$ = 192 and 205 GeV in our model \cite{nls10}.
In Fig.\ \ref{nlsusy3} the contours of $m_{S_1}$ and those of $(\sigma_1+\sigma_2)$
are plotted.
We find $(\sigma_1+\sigma_2) \le$ 50 fb and
$(\sigma_1+\sigma_2)_{{\rm min}} \approx$ 26 fb in the hatched region.
This region contains part of the $m_{S_1}$ = 0 contour, which means that
LEP 2 with $\sqrt{s}$ = 175 GeV would not be able to put any
constraints on $m_{S_1}$.

In order to see whether LEP 2 can put a constraint on the quartic coupling
constant $\lambda_0$, we scan the parameter space $|k| \le$ 0.7,
$|\lambda_1| \le$ 1, 0 $< \lambda_2 <$ 1, 2 $< \tan \beta <$ 15,
and 150 GeV $< m_C <$ 1000 GeV and plot $(\sigma_1 + \sigma_2)$ as a
function of $\lambda_0$, for $\sqrt{s}$ = 175 GeV in Fig.\ \ref{nlsusy4}.
About 10$^6$ points are considered.
Again one sees that with a discovery limit of 30 fb $\sim$ 50 fb LEP 2 with
175 GeV will not be able to put any experimental limit on $\lambda_0$, either.

As expected, the situations both with $\sqrt{s}$ = 192 GeV and
with $\sqrt{s}$ = 205 GeV are much more favorable.
We scan the same parameter space as that of Fig.\ \ref{nlsusy4}a and
determine $(\sigma_1 + \sigma_2)$ as a function of $\lambda_0$ at these c.m.
energies.
We then plot the results in Fig.\ \ref{nlsusy4}b for $\sqrt{s}$ = 192 GeV and in Fig.\ \ref{nlsusy4}c
for 205 GeV.
Fig.\ \ref{nlsusy4}b shows that $(\sigma_1 + \sigma_2)$ is greater than 50 fb
for $\lambda_0 \le$ 0.54.
Thus LEP 2 with $\sqrt{s}$ = 192 GeV would be able to put an experimental
lower limit on $\lambda_0$ as
\begin{equation}
 \lambda_{0,{\rm EXP}} \ge 0.53 \ .
\end{equation}
This experimental lower limit implies via Eq.\ (\ref{ms1bound}) an experimental lower
limit on the upper limit on $m_{S_1}$ as
\begin{equation}
m_{S_1,{\rm max},{\rm EXP}} \ge 1.92 \cdot \lambda_{0,{\rm EXP}} \cdot m_Z
 \approx 92 \ {\rm GeV}  \ .
\end{equation}
%
\begin{figure}
\epsfxsize=6.5cm
\mbox{\epsffile{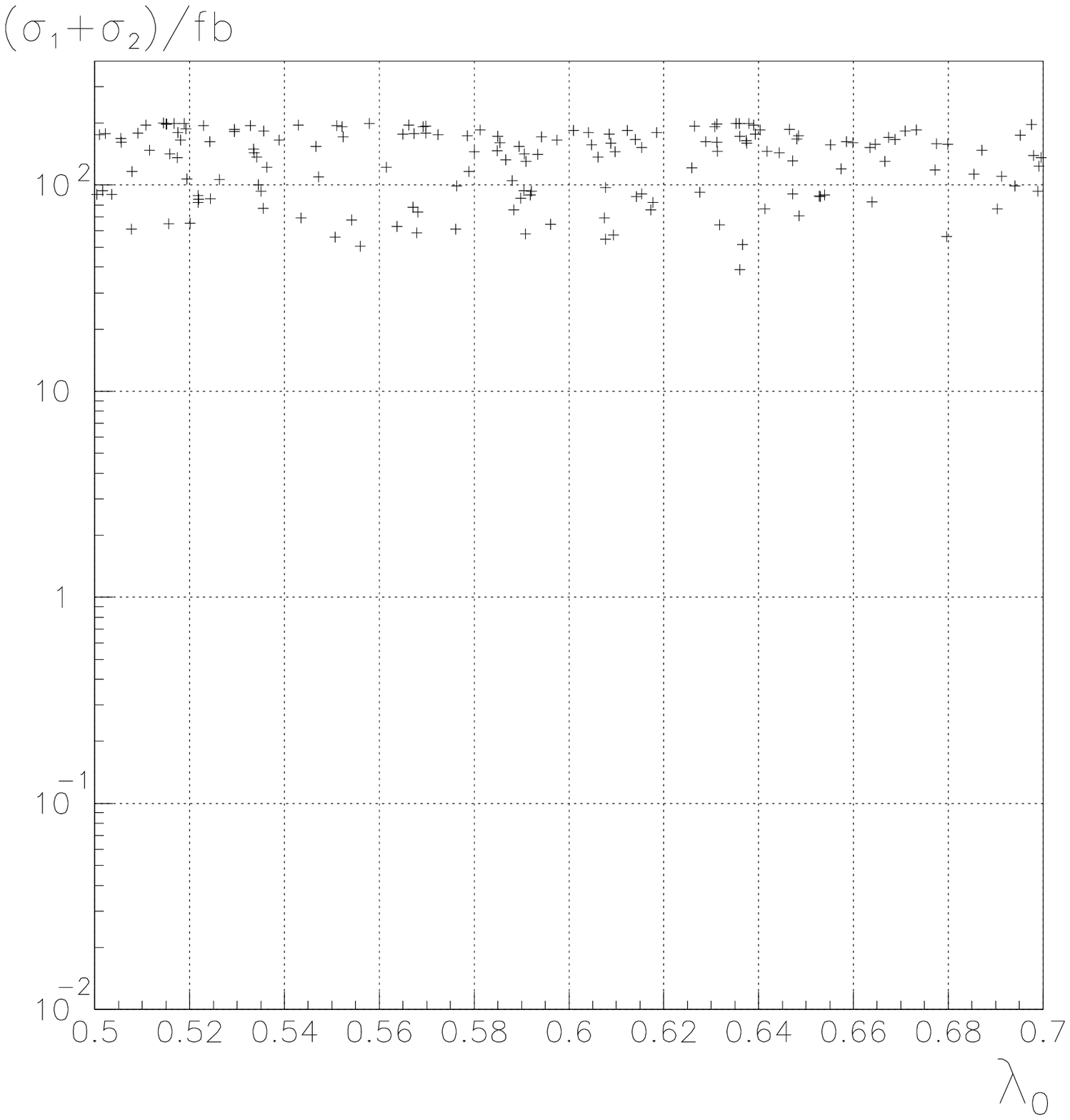}}
\hspace*{\fill} \epsfxsize=6.5cm
\mbox{\epsffile{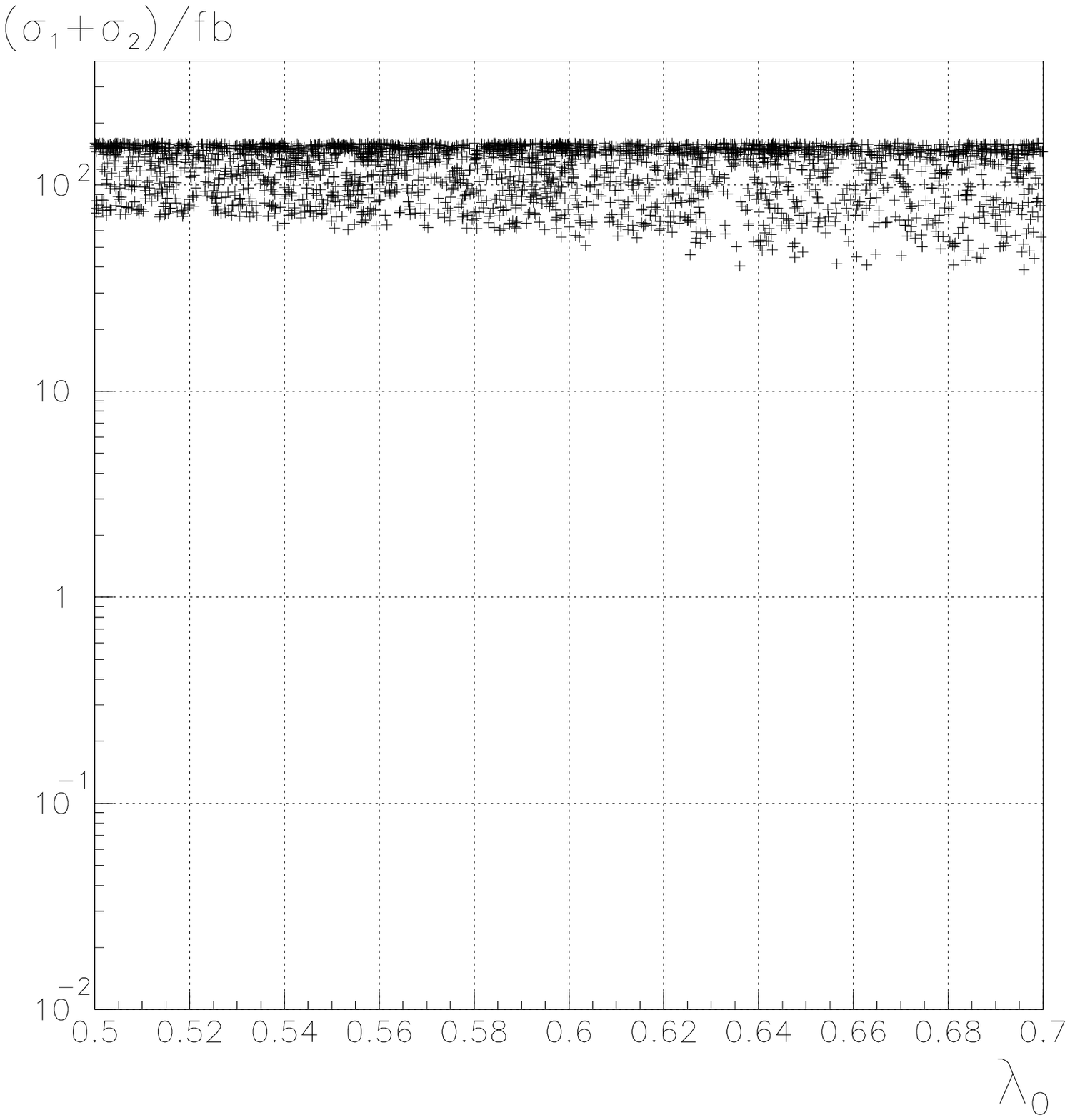}}
\\[-2cm] \hspace*{5.35cm} \large (a)
\\[-\baselineskip] \hspace*{13.45cm} \large (b)
\\[2cm]\\[-3\baselineskip]
\caption[plot]{(a) The cross section $(\sigma_1 + \sigma_2)$ with
$m_{S_1} \le$ 10 GeV at $\sqrt{s}$ = 192 GeV, as function of $\lambda_0$, for
$| k | \le$ 0.7, $| \lambda_1 | \le$ 1, 0 $< \lambda_2 <$ 1,
2 $< \tan \beta <$ 15, and 150 GeV $< m_C <$ 1000 GeV.\ \ ---\ \ (b)
The same as (a), except for $\sqrt{s}$ = 205 GeV
and $m_{S_1} \le$ 27 GeV.}
\label{nlsusy5}
\end{figure}
Similarly from Fig.\ \ref{nlsusy4}c we conclude for $\sqrt{s}$ = 205 GeV
\begin{equation}
 \lambda_{0,{\rm EXP}} \ge 0.61
\end{equation}
and
\begin{equation}
m_{S_1,{\rm max},{\rm EXP}} \ge 107 \ {\rm GeV} \ .
\end{equation}

Now, we turn to the question of imposing a lower limit on $m_{S_1}$ itself.
In Fig.\ \ref{nlsusy5}a we plot $(\sigma_1 + \sigma_2)$ for $\sqrt{s}$ = 192 GeV
with the constraint $m_{S_1} \le$ 10 GeV.
We omit the points for $\lambda_0 <$ 0.5 as in the region
$(\sigma_1 + \sigma_2) \ge$ 50 fb.
For $m_{S_1} \ge$ 10 GeV + 1 GeV = 11 GeV, there are points with
$(\sigma_1 + \sigma_2) <$ 30 fb.
We find that $(\sigma_1 + \sigma_2)$ for $m_{S_1} <$ 10 GeV is always greater
than 30 fb.
This implies that LEP 2 with $\sqrt{s}$ = 192 GeV and discovery limit 30 fb
would be able to put an experimental lower limit on $m_{S_1}$ as
\begin{equation}
m_{S_1,{\rm EXP}} \ge 10 \ {\rm GeV} \ .
\end{equation}
In Fig.\ \ref{nlsusy5}b we plot the same as in 5a for $\sqrt{s}$ = 205 GeV.
In this case we obtain
\begin{equation}
m_{S_1,{\rm EXP}} \ge 27 \ {\rm GeV} \ .
\end{equation}

\section{Higgs Production at LC 500, 1000, and 2000}

As discussed in section \ref{nlsmodel}, the upper bound of $m_{S_1}$ is about 130 GeV.
Thus, if the collider energy $\sqrt{s}$ is larger than
${\rm E}_C = m_Z$ + 130 GeV, which is a kind of threshold energy,
the production via the Higgs-strahlung is possible in the whole
parameter space for at least one of $S_i$ $(i=1,2,3)$.
In this case one should consider the productions of $S_1, S_2$, and $S_3$
simultaneously.
In order to be systematic, we consider the production cross sections
of $S_1, S_2$, and $S_3$ via the Higgs-strahlung process, denoted respectively
by $\sigma_1$, $\sigma_2$, and $\sigma_3$:
\begin{eqnarray}
   \sigma_1 (m_{S_1}) &=& \sigma_{{\rm SM}}(m_{S_1}) R^2_1  \cr
   \sigma_2 (m_{S_2}) &=& \sigma_{{\rm SM}}(m_{S_2}) R^2_2  \\
   \sigma_3 (m_{S_3}) &=& \sigma_{{\rm SM}}(m_{S_3}) (1 -R^2_1 -R^2_2) \ , \nonumber
\end{eqnarray}
where $\sigma_{{\rm SM}}(m)$ is the cross section in the standard model
for the production of the Higgs boson of mass $m$ via the Higgs-strahlung 
process.

A useful observation is that $\sigma_i (m_{{S_i},{\rm max}}) \le
\sigma_i (m_{S_i})$, which allows to derive the parameter-independent
lower limit of $\sigma_i$ as we will show in the following.
At first, we determine the cross sections $\sigma_1 (R_1, R_2, m_{S_1})$,
$\sigma_2 (R_1, R_2, m_{S_2,{\rm max}})$, and
$\sigma_3 (R_1, R_2, m_{S_3,{\rm max}})$ at a fixed set of $m_{S_1}$, $R_1$,
and $R_2$.
Secondly, we keep $R_1$ and $R_2$ fixed, while varying $m_{S_1}$ from its
minimum to maximum and determine the quantity
\begin{equation}
\sigma(R_1, R_2) = {\rm min} [{\rm max}(\sigma_1, \sigma_2, \sigma_3)]
\  \  \  \  \   (0 \le m_{S_1} \le m_{S_1,{\rm max}}) \\ \nonumber
\label{sigmar1r2}
\end{equation}
where $\sigma_1 = \sigma_1(R_1, R_2, m_{S_1})$ and
$\sigma_i = \sigma_i(R_1, R_2, m_{S_i,{\rm max}})$  $(i = 2,3)$.

As last step, we vary $R_1^2$ and $R_2^2$ from 0 to 1
$(0 \le R_1^2 \le 1$ and $0 \le R_2^2 \le 1)$ and plot $\sigma(R_1, R_2)$
in the $R_1^2$-$R_2^2$ plane.
%
\begin{figure}
\epsfxsize=6.5cm
\mbox{\epsffile{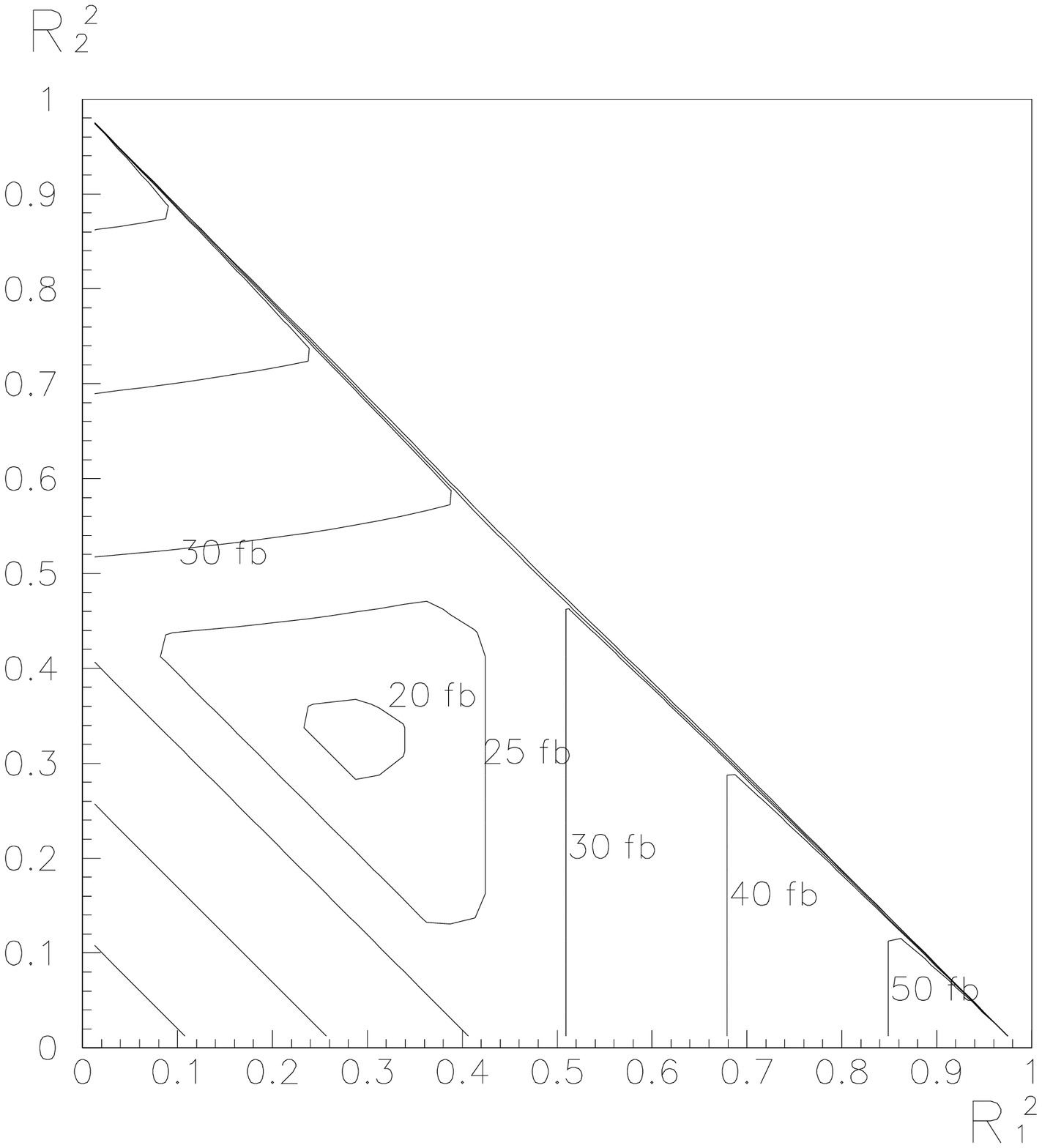}}
\hspace*{\fill} \epsfxsize=6.5cm
\mbox{\epsffile{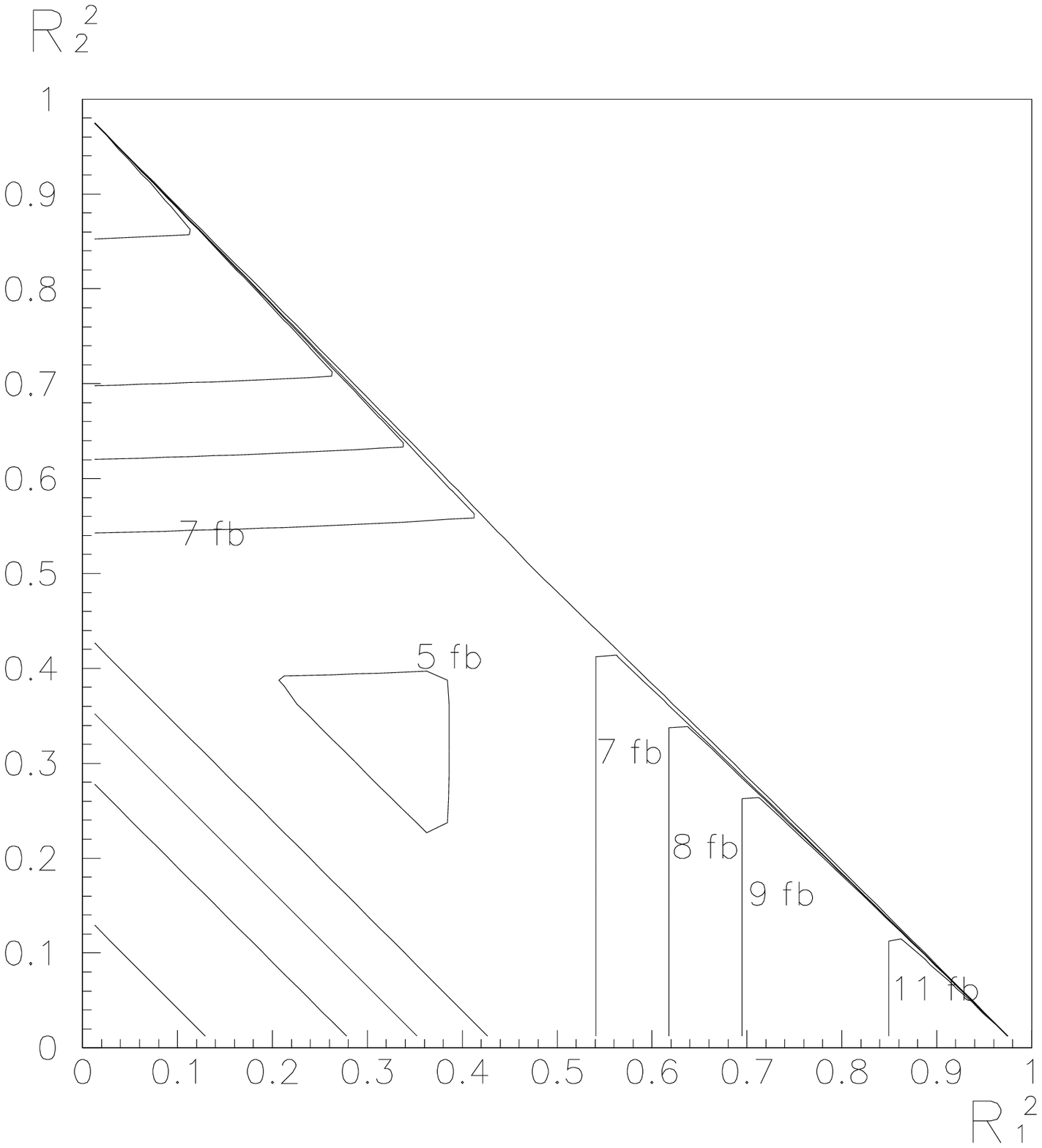}}
\\[-3cm] \hspace*{5.45cm} \large (a)
\\[-\baselineskip] \hspace*{13.45cm} \large (b)
\\[3cm] \epsfxsize=6.5cm \hspace*{3.85cm} 
\mbox{\epsffile{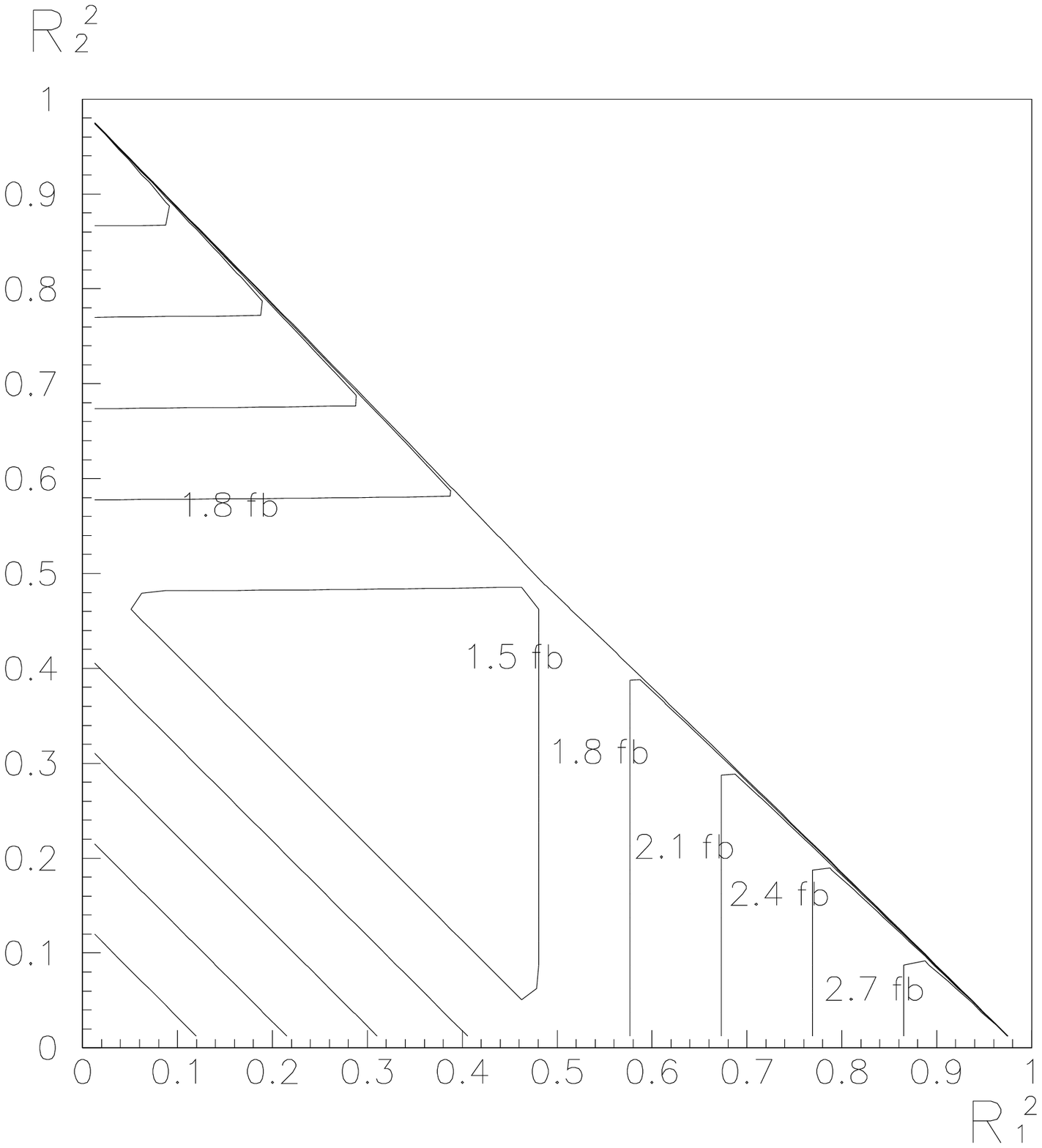}}
\\[-3cm] \hspace*{9.3cm} \large (c)
\\[3cm]\\[-2\baselineskip]
\caption[plot]{
Contour lines of $\sigma(R_1, R_2)$ as defined by Eq.\ (\ref{sigmar1r2}) for
(a) $\sqrt{s} = 500$ GeV,
(b) $\sqrt{s} = 1000$ GeV,
(c) $\sqrt{s} = 2000$ GeV. The minimum values are 19 fb for (a), 5 fb for (b), and 1.2 fb for (c).}
\label{nlsusy6}
\end{figure}
For $\sqrt{s} \le {\rm E_T} = m_Z + m_{S_1,{\rm max}} \approx 222$ GeV
this plot produces null results because $\sigma(R_1, R_2)$ = 0, which is the
case for LEP 2.
For $\sqrt{s} > {\rm E_T}$, $\sigma(R_1, R_2)$ never vanishes in the entire
$R_1^2$-$R_2^2$ plane and the minimum value of $\sigma(R_1, R_2)$ is a
parameter-independent lower limit of one of $\sigma_1, \sigma_2, \sigma_3$.
Thus, this minimum is a characteristic quantity of the model.

In Fig.\ \ref{nlsusy6}a we plot $\sigma(R_1, R_2)$ for $\sqrt{s}$ = 500 GeV.
The minimum of $\sigma(R_1, R_2)$ in the plane is about 19 fb,
which means that one of $S_i$ will be produced with $\sigma_i \ge$ 19 fb
for $\sqrt{s}$ = 500 GeV.
For a discovery limit of 50 events per year one would need an integrated
luminosity of about 2.5 fb$^{-1}$, which is a realistic one.

Fig.\ \ref{nlsusy6}b and \ref{nlsusy6}c show $\sigma(R_1, R_2)$ for $\sqrt{s}$ = 1000 GeV and
2000 GeV, respectively.
The minimum $\sigma(R_1, R_2)$ is about 5 fb for $\sqrt{s}$ = 1000 GeV and
1.2 fb for $\sqrt{s}$ = 2000 GeV.
The conclusion is that this model may most probably be tested at future LC 500,
1000, and 2000 colliders.

\newpage

\section{Qualitative Discussion about Decay Modes}

The main purpose of the present paper is to investigate at which energy and luminosity our model could in principle be tested.  This is what we have just done considering the production of on-shell $S_i$.  However, for experimental searches, more detailed informations are needed, in particular on their decay modes.  Comprehensive investigations in this respect are under way, similar to the investigations done for the MSSM \cite{nls11}.  Here, we merely make a few quantitative remarks.  

The dominant decay modes of $S_1$ are those into $b$ quark and
$\tau$ lepton pairs, except for the case where $m_{S_1}$ approaches its maximum value.  In this case, $S_1$ behaves like the standard model Higgs boson, and other decay modes, for example those into pairs of gauge bosons will become important, with partial widths that could become comparable to those of the $b \bar b$ channel for large $\tan\beta$.  An important signature of $S_1$ is certainly its upper mass bound of about 130 GeV.

The decay modes of the heavy bosons $S_2$, $S_3$ could be more complex,
depending on $\tan\beta$.  For large $\tan\beta$ these bosons decay dominantly to $b\bar b$ and $\tau^+ \tau^-$.  In the MSSM, the decay of the heavy neutral scalar Higgs boson into a pair of light scalar or pseudoscalar bosons can be dominant in the parameter region where the mass of the heavy Higgs boson approaches its maximum \cite{nls11}.  The question whether this could happen in our model, too, is under investigation.  For small $\tan\beta$, the decay modes into pairs of
light Higgs bosons, gauge boson pairs, and mixed pairs of Higgs and gauge bosons will become important.  Above the $t \bar t$ threshold, $S_3$ will decay dominantly into $t$ quark pairs.  The upper bound of $m_{S_2}$ is smaller than the
threshold.  We numerically determined bounds for the masses of $S_2$, $S_3$, $P_1$, and $P_2$ by systematically scanning the parameter space and obtained
$\mbox{55 GeV}\ {<\atop\sim}\ m_{S_2}\ {<\atop\sim}\ \mbox{260 GeV}$,
$m_{S_3}\ {>\atop\sim}\ \mbox{150 GeV}$, $m_{P_1}\ {<\atop\sim}\ \mbox{240 GeV}$,
and $m_{P_2}\ {>\atop\sim}\ \mbox{120 GeV}$

Another interesting question is how to distinguish the Higgs sector of our model from those of other models, in particular, from that of the NMSSM, which 
has the same Higgs particle spectrum.  The Higgs sectors should be easy to 
distinguish if some of the s-particles of the
NMSSM were light enough for the Higgs bosons to decay into.  Otherwise, the decay patterns should be very similar in both models.  A theoretical possibility
to distinguish the models arises from the number of free parameters of the 
Higgs sector.  Although both models have the same number of parameters on tree level, the numbers differ on loop level.  For the NMSSM, the number increases
due to the contributions of the s-particles, whereas for our model, it remains
the same, i.e. six.  So once all Higgs bosons were found, our model could be determined completely by six independent experiments.

\section{Conclusion}

We demonstrated that at LEP 2 with $\sqrt{s}$ = 175 GeV no bounds on
$m_{S_1}$ and $\lambda_0$ can be derived, whereas LEP 2 with
$\sqrt{s}$ = 192 GeV and $\sqrt{s}$ = 205 GeV will be able to put experimental
lower bounds on $\lambda_0$, $m_{S_1,{\rm max}}$, and $m_{S_1}$.
Our analysis predicts
\begin{eqnarray}
\lambda_{0,{\rm EXP}} & \ge & 0.53 \ (0.61)                    \cr
m_{S_1,{\rm max},{\rm EXP}} & > & 92 \ {\rm GeV} \ (107 \ {\rm GeV})  \\
m_{S_1,{\rm EXP}} & > & 10 \ {\rm GeV} \ (27 \ {\rm GeV})    \nonumber
\end{eqnarray}
for $\sqrt{s}$ = 192 GeV (205 GeV).
We also derived a lower limit of the production cross sections of the scalar
Higgs bosons to be 19 fb, 5 fb and 1.2 fb at $e^+e^-$ colliders with
$\sqrt{s}$ = 500 GeV, 1000 GeV, and 2000 GeV respectively,
which are large enough to test the model conclusively.

One should remark that the above results are based on tree level
calculations.


\vfil \eject

\end{document}